**Gizachew Tiruneh, Ph. D., Department of Political Science, University of Central Arkansas, Conway, Arkansas**

Most professional golfers and analysts think that winning on the PGA Tour peaks when golfers are in their thirties. Rather than relying on educated guesses, we can actually use available statistical data to determine the actual ages at which golfers peak their golf game. We can also test the hypothesis that age affects winning professional golf tournaments. Using data available from the Golf Channel, the PGA Tour, and LPGA Tour, I calculated and provided the mean, the median, and the mode ages at which professional golfers on the PGA, European PGA, Champions, and LPGA Tours had won over a five-year period. More specifically, the ages at which golfers on the PGA, European PGA, Champions Tour, and LPGA Tours peak their wins are 35, 30, 52, and 25, respectively. The regression analyses I conducted seem to support my hypothesis that age affects winning professional golf tournaments.

**Introduction**

Aging tends to decrease the flexibility, bone mass, and strength of the human body, leading to a decline in the level of physical activity of the average person (Birrer, 1989).[1] Even professional athletes are not immune from the decline of talent due to aging. In swimming, cycling, and weightlifting, for instance, most athletes reach their prime in their 20s and early 30s (Wilmore, David, Costill, & Kenney, 2008). How about the age at which professional golfers peak? The conventional wisdom about professional golfers is that they go through three consecutive phases; first, they go through the learning process, acquiring both physical and mental skills of the game. Second, after a few years of experience, they reach their prime. Lastly, due largely to physiological factors, their skills start to decline (Berry & Larkey, 1999). There is only one identifiable study conducted to investigate such a relationship, however. Based on 489 golfers and using scoring average as the measure of performance over the course of several years, Berry et al. (1999) found that most golfers, who played in the four Major championships (the Masters, U.S. Open, British Open, and the PGA), peak on average between 30 and 35 years.[2] They, however, did not include non-Major golf tournaments in their analysis. They also did not deal with other major tours, such as the European Professional Golf Association (European PGA), the Champions, and the Ladies Professional Golf Association (LPGA) Tours. Nor did they measure golfers' performance by number of wins.

---

[1] Morgan et al. (1999) also found that younger amateur golfers' bodies tend to be more flexible when playing golf than those of adult and senior amateur golfers'.

This paper uses winning, as opposed to scoring averages, as a measure of performance and investigates the ages at which professional golfers peak their wins in both Major and non-Major golf tournaments for the years between 2003 and 2007. The analysis also covers four major golf tours. Specifically, using data available from the Golf Channel, the Professional Golf Association (PGA) Tour, the European PGA Tour, and the LPGA Tour, I calculated and provided the mean, the median, and the mode ages at which professional golfers on these tours had won over a five-year period. I have found that the ages at which golfers peak their wins on the PGA, European PGA, Champions, and LPGA Tours are 35, 30, 52, and 25, respectively. Moreover, the regression analyses I conducted seem to support my hypothesis that age affects winning professional golf tournaments.

---

[2] Lockwood (1999) studied the effect of aging on amateur golfers and found that amateur golfers peak their golfing skills between the ages of 20 and 39.

## The PGA Tour

I relied on 239 major tournaments in which PGA Tour golfers played between 2003 and 2007 to calculate the mean, the median, and the mode ages of winners. Table 1 and Figure 1a through 1d show these results. Model 1, in Table 1, shows that the mean or average and the median ages of winning PGA tournaments were 35.05 and 35 years, respectively. The age at which PGA Tour players' wins peaked (the mode) was 31. Fred Funk, at 51, was the oldest winner on the PGA Tour between 2003 and 2007; he won the Mayakoba Classic tournament in 2007.

It has been known for a while, however, that Tiger Woods is one of the best (if not the best) golfers to have played the game of golf. He had won 65 tournaments so far, and 27 of his wins came between 2003 and 2007. Because of this even Tiger Woods' peers like Ernie Els think that he is well above everybody else with respect to his golf skills. In other words, if one was to analyze the skills of professional golfers, Tiger Woods' extraordinary talent would make him a deviant or an outlier case. In Model 2, I excluded Tiger Woods from the analysis. The mean, the median, and the mode ages of winning PGA tours were 35.65, 35, and 35, respectively. The main difference between the analysis in Model 1 and that in Model 2 is that in the absence of Tiger Woods, the mode or the age at which tour players peak their wins increased from 31 to 35.

Another golfer who may be considered an outlier due to several wins in his forties is Vijay Singh. Singh won 23 tournaments between 2003 and 2007, after he became 40 years old. When Vijay Singh is excluded from the data, in Model 3, the mean, the median, and the mode ages became 34.36, 34, and 31 respectively. The main change from Model 1 is that the mean and the median ages became smaller by about a year. And the main change from Model 2 is that the mode age decreased from 35 to 31, which seemed to be influenced by the presence of Tiger Woods. In Model 4, I excluded both Tiger Woods and Vijay Singh from the analysis. Interestingly, the mean, the median, and the mode ages of winning PGA tournaments became almost equal, 35 years old. When these measures are equal, the distribution of the data is said to approximate or take the form of a bell curve. Thus, we can say that about 68 % of winners were between one standard deviation from the mean. Since the standard deviation is 6.15, we can say that about 68 % of the winners were between the ages of 29 and 41. Similarly, about 95 % of the winners were within two standard deviations from each side of the curve. That is, about 95 % of the winners were between the ages of 23 and 47. It is, thus, safe to say that under normal circumstances (that is, without the presence of outlier golfers like Woods and Singh), the age at which the PGA Tour golfers peak their wins is 35. It is interesting to note that perhaps because Berry et al. (1999) relied only on the four Major championships (as opposed to all tournaments) and on golfers' scoring averages (as opposed to winning tournaments), their peak years (between 30 and 35) and mine (35) are not identical.

**Table 1: Mean, Median, and Mode Ages of Winning PGA Tour (2003-2007)**

|  | Model 1 Singh & Woods Included | Model 2 Woods Excluded | Model 3 Singh Excluded | Model 4 Singh & Woods Excluded |
|---|---|---|---|---|
| Mean | 35.05 | 35.65 | 34.36 | 34.95 |
| Median | 35.00 | 35.00 | 34.00 | 35.00 |
| Mode | 31.00 | 35.00 | 31.00 | 35.00 |

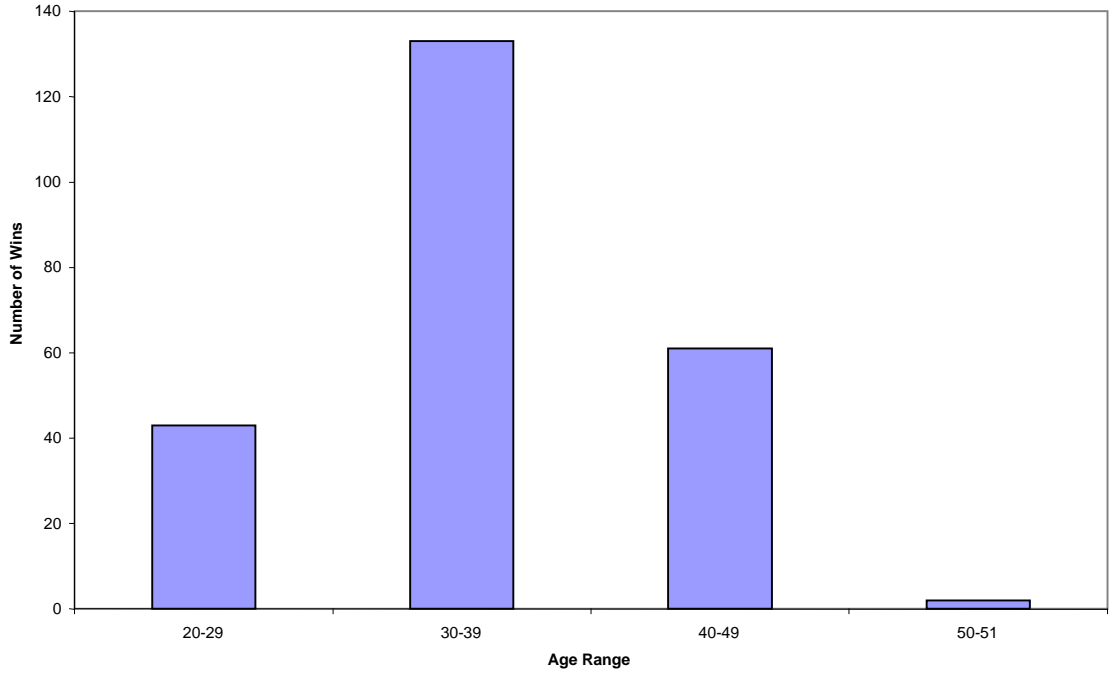

**Fig. 1a: PGA Tour Wins (2003-2007, both T. Woods and V.Singh included)**

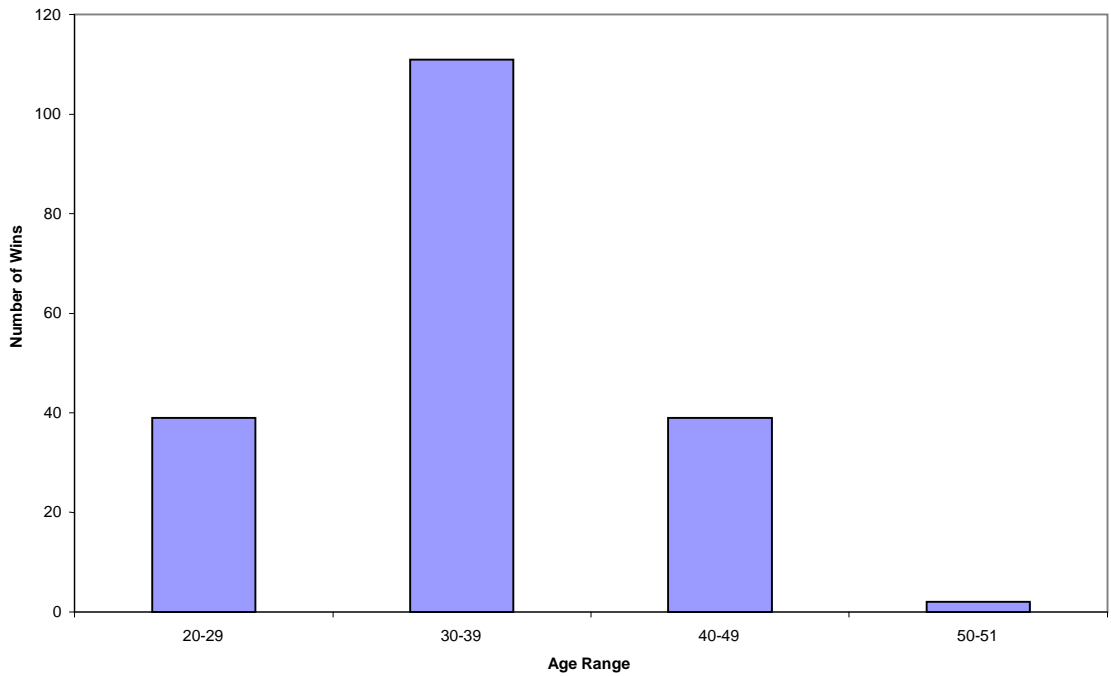

**Fig. 1b: PGA Tour Wins (2003-2007, T. Woods & V. Singh excluded)**

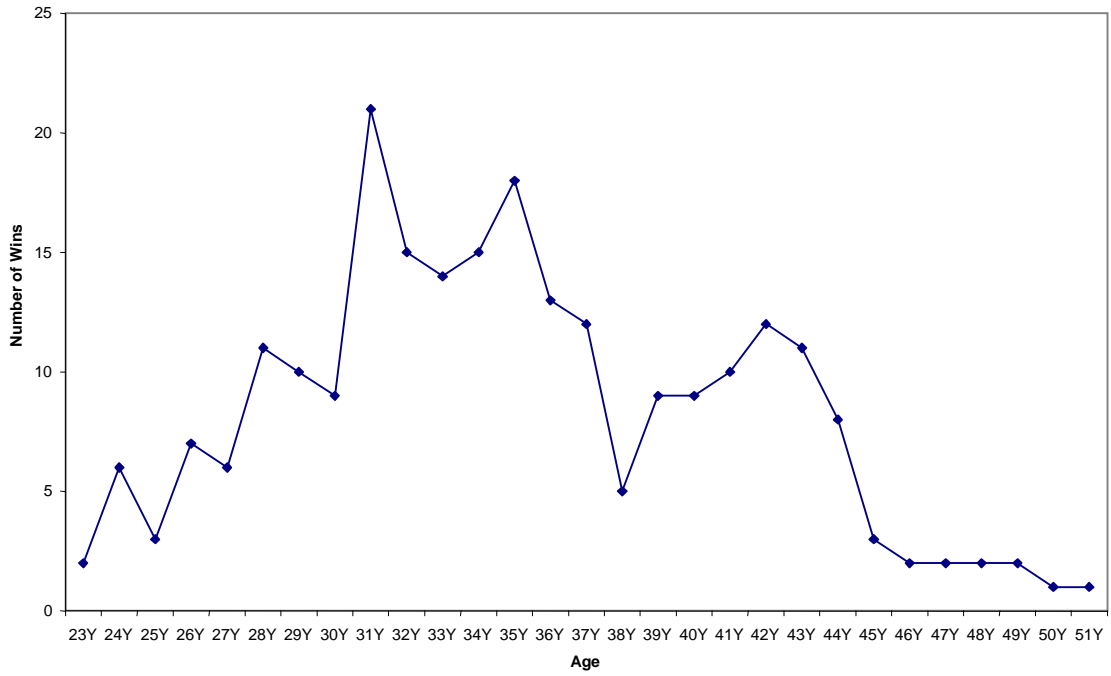

**Fig. 1c: PGA Tour Winning Trend (2003-2007, T. Woods & V. Singh included)**

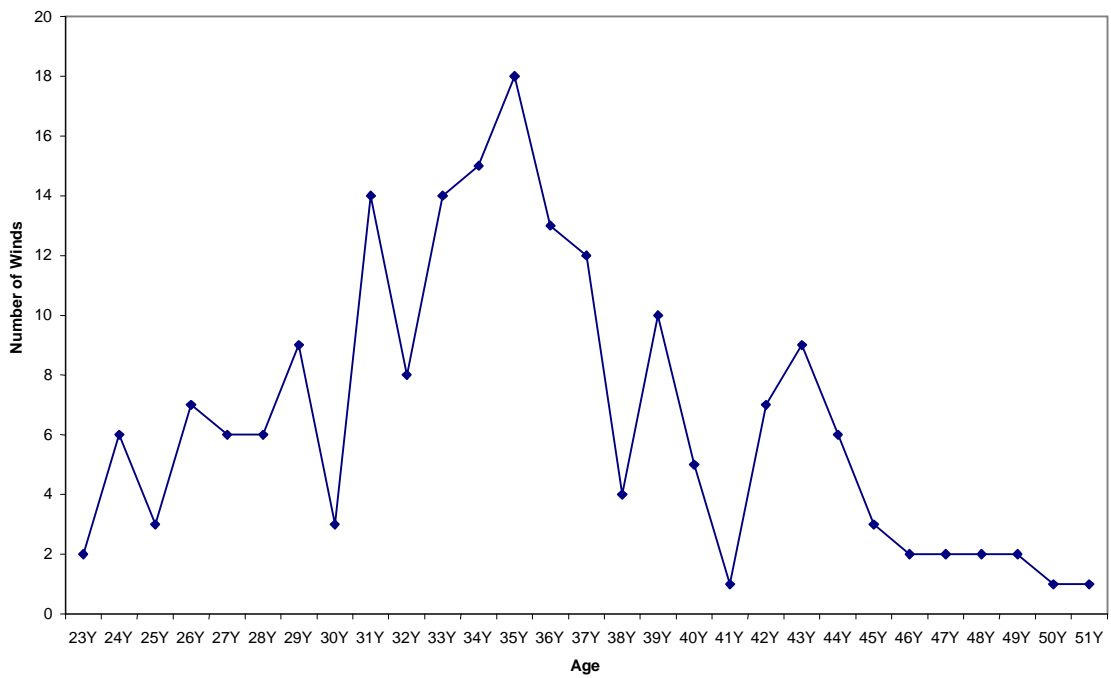

**Fig. 1d: PGA Tour Winning Trend (2003-2007, T. Woods & V. Singh excluded)**

## The European PGA Tour

There were 237 major tournaments in which European PGA Tour players played between 2003 and 2007. These tournaments included the World Golf Championship (WGC) events and the four majors. The mean, the median, and the mode ages of the winners were 32.50, 32, and 30, respectively. These statistics are shown in Table 2 and Figure 2a and 2b. European Tour players seem to peak their wins when they are 30 years old, which is about 5 years lower than U.S. PGA Tour players. The mean and the median ages of winning were also smaller by about 3 years. More golfers in the European Tour also had won in their twenties than in their forties. Mark O'Meara, the American, was the oldest winner on of the European PGA Tour between 2003 and 2007; he won the Dubai Classic tournament at the age of 47 in 2004.

**Table 2: Mean, Median, and Mode Ages of Winning in the European Tour (2003-7)**

| **Mean** | 32.50 |
|---|---|
| **Median** | 32.00 |
| **Mode** | 30.00 |

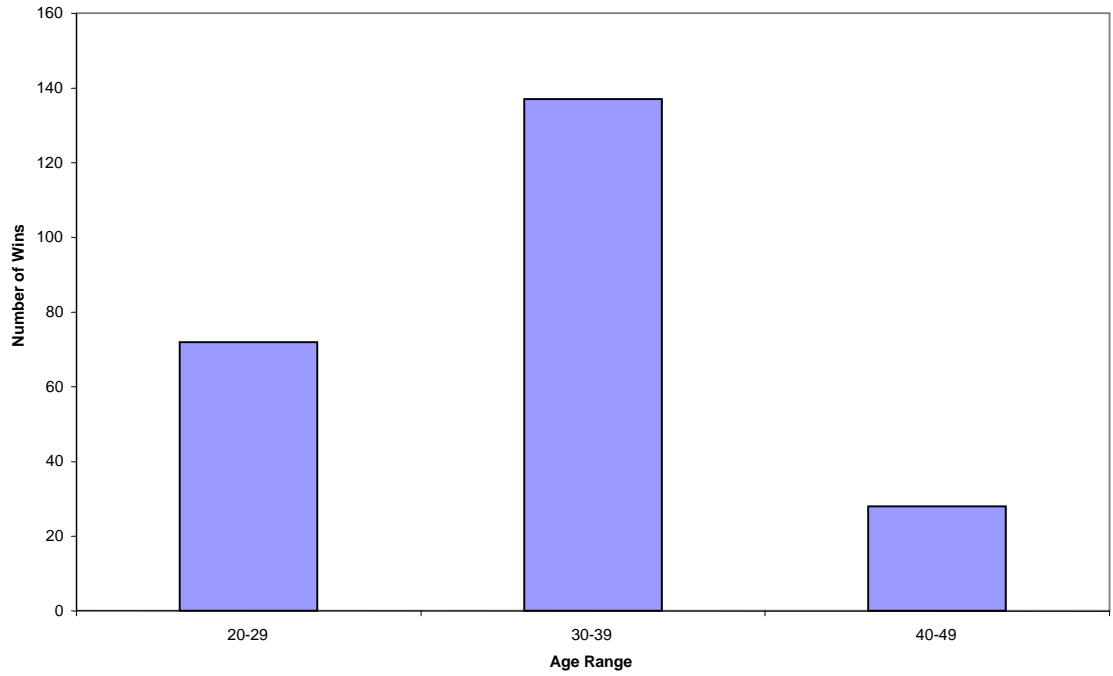

**Fig. 2a: European Tour Wins (2003-2007)**

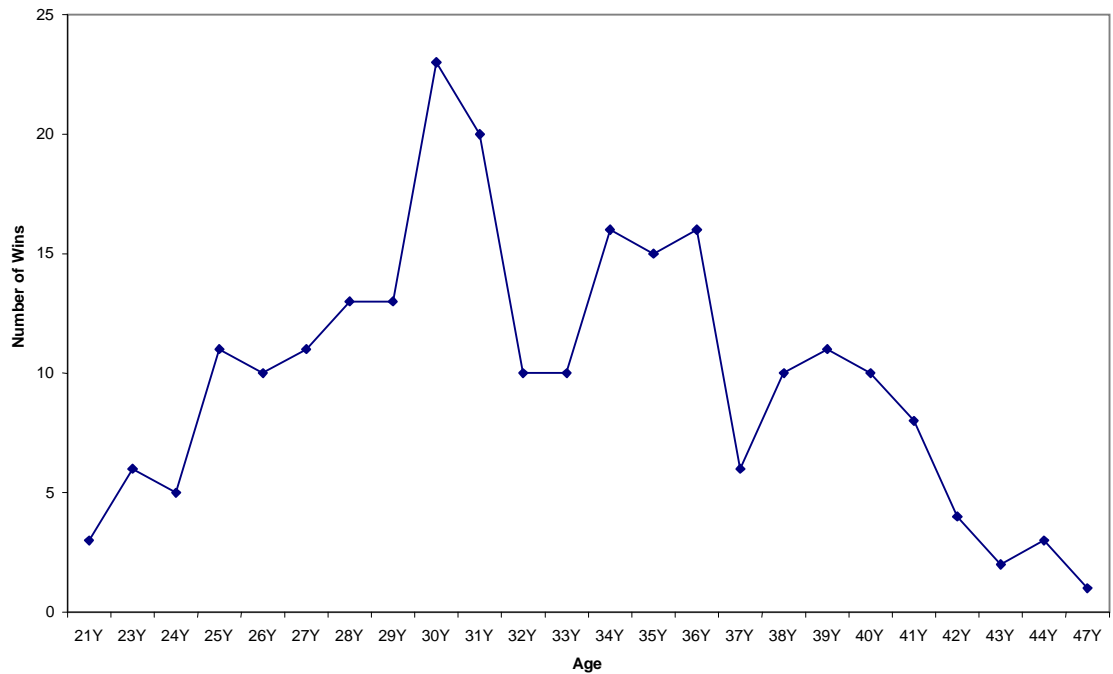

**Fig. 2b: European Tour Winning Trend (2003-2007)**

## The Champions Tour

There were 146 tournaments in which Champions Tour players played between 2003 and 2007. The mean, the median, and the mode ages of the winners were 54.21, 54, and 52, respectively. These statistics are shown in Table 3 and Figure 3a and 3b. Champion Tour golfers seem to peak their wins when they are 52 years old. Not surprisingly, winning in this Tour started to decline very quickly as golfers were aging. Hale Irwin was the oldest winner on the tour between 2003 and 2007; he won the MasterCard Championship in 2007 at the age of 62.

**Table 3: Mean, Median, and Mode Ages of Winning in the Champion Tour (2003-7)**

| **Mean** | 54.21 |
|---|---|
| **Median** | 54.00 |
| **Mode** | 52.00 |

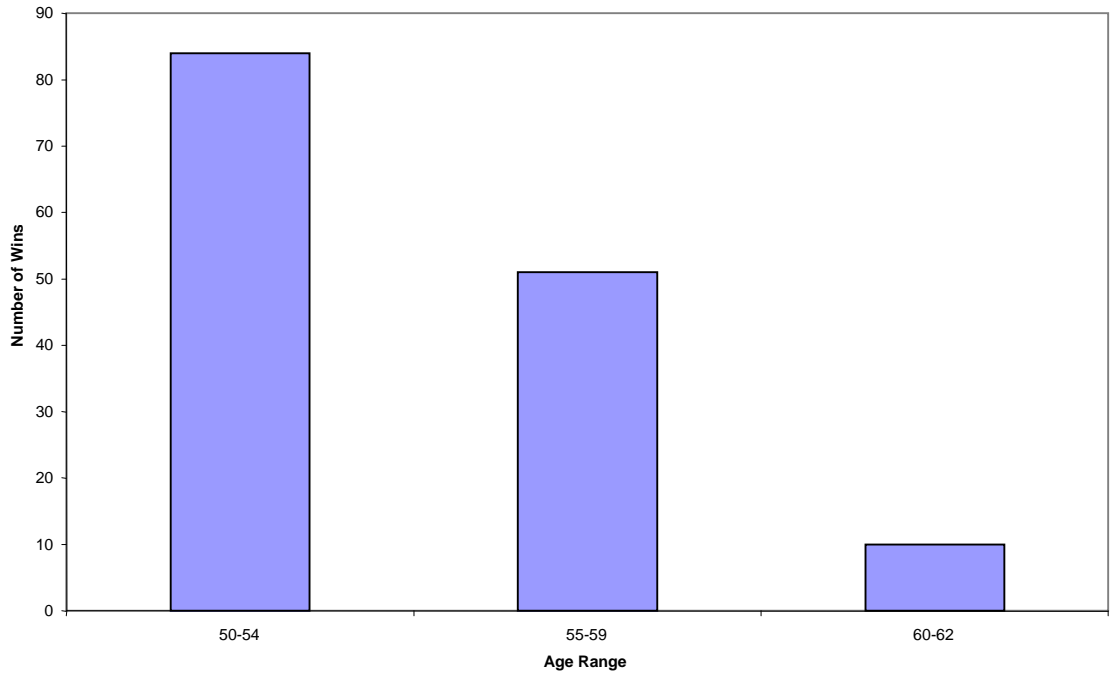

**Fig. 3a: Champion Tour Wins (2003-2007)**

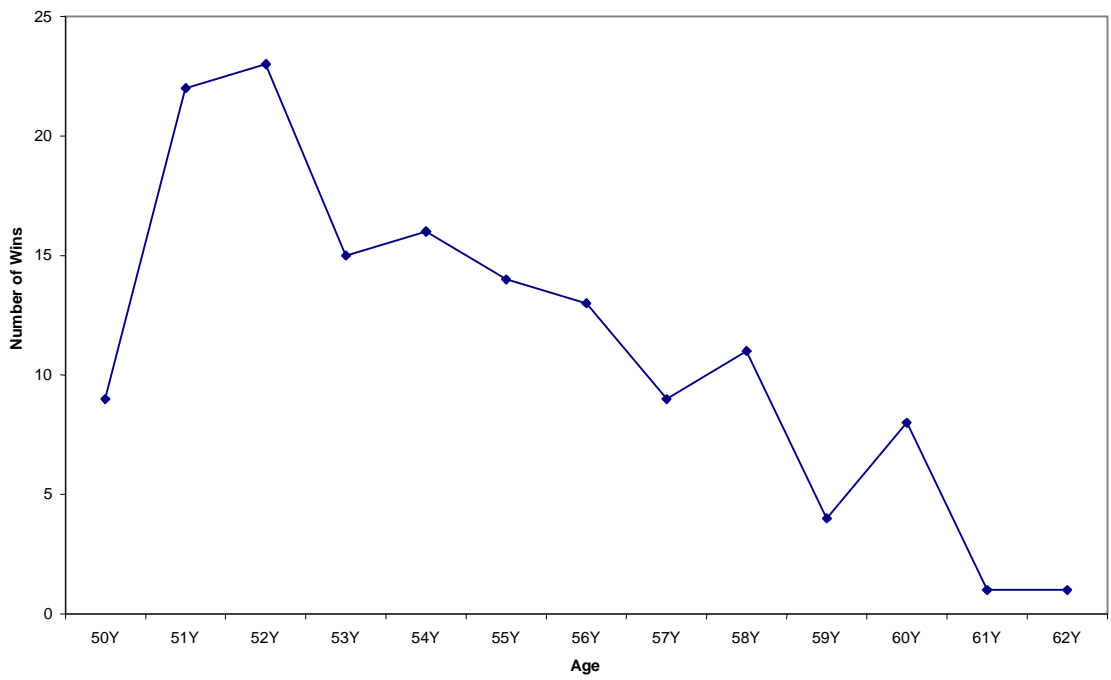

**Fig. 3b: Champions Tour Winning Trend (2003-2007)**

**The LPGA Tour**

There were 159 tournaments played by the LPGA Tour golfers between 2002 and 2006. Because the data for 2007 were not available on the LPGA Tour web site, I included the 2002 results in order to make the analysis expand over 5 years. These results are summarized and presented in Table 4 and Figure 4a through 4d. As shown in Model 1, the mean and the median ages of winning for women golfers were 29.90 and 30, respectively. There were two mode years, 25 and 32, however, making the distribution of wins in the LPGA Tour bimodal (see Fig. 1c).

We also know that Annika Sorenstam has been a dominant (or outlier) player on the LPGA Tour. Indeed, the reason we have a second mode, 32, in Model 1 is mainly because of Sorenstam's 11 wins in 2002 at age 32. She had won over 70 tournaments, and 38 of them came between 2002 and 2006. Given the foregoing, I show in Model 2 the results I obtained after I excluded Sorenstam from the analysis. The mean, the median, and the mode ages of winning in this tour became 28.71, 27, and 25, respectively. Interestingly, the distribution of winning became unimodal, and the age at which the LPGA Tour players peaked their wins remained 25 even when Sorenstam is excluded from the analysis. The mode age, 25, which is perhaps the norm without outlier golfers like Sorenstam, is 5 and 10 years lower than those for the European and U.S. PGA Tour players, respectively. Because Sorenstam was in her thirties between 2002 and 2006, the mean and the median ages of winning for the rest of women golfers were lower by about 1 and 3 years, respectively. Beth Daniel, at 47, was the oldest player to win on the LPGA Tour between 2002 and 2006; she won the BMO Financial Group Canadian Women's Open in 2003.

**Table 4: Mean, Median, and Mode Ages of Winning in the LPGA Tour (2002-2006)**

|  | Model 1 | Model 2 |
|---|---|---|
|  | Sorenstam Included | Sorenstam Excluded |
| Mean | 29.90 | 28.71 |
| Median | 30.00 | 27.00 |
| Mode | 25.00, 32.00 * | 25.00 |

*Note: In strict statistical sense, we should also have two means and medians in a bimodal distribution.

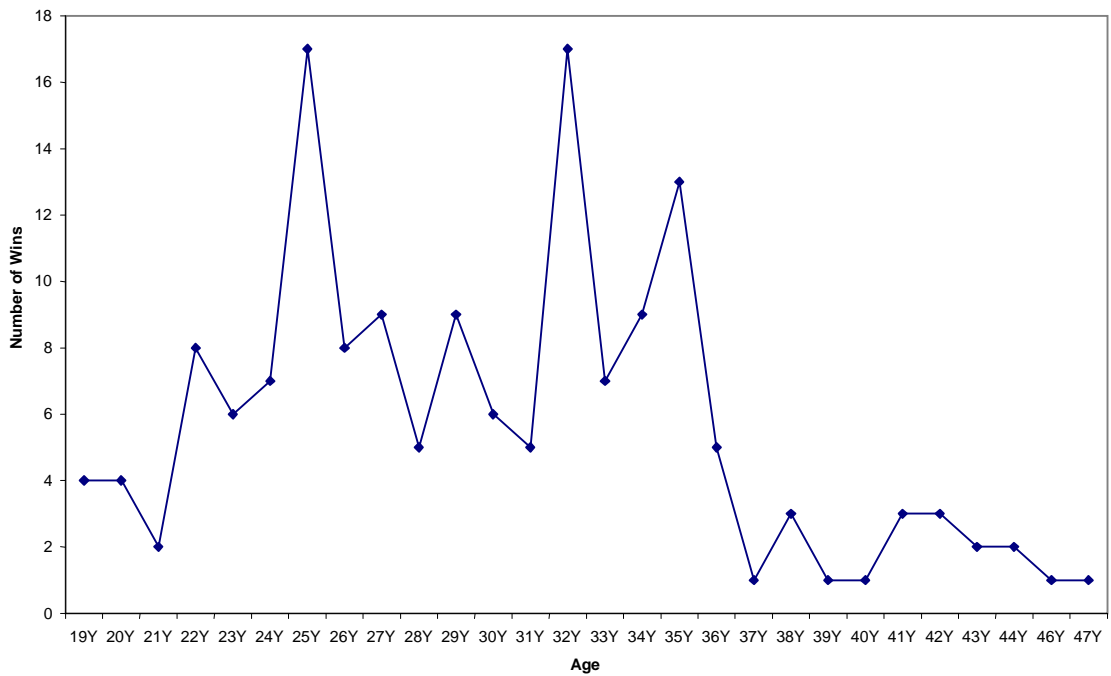

**Fig. 4c: LPGA Tour Winning Trend (2002-2006, A. Sorenstam included)**

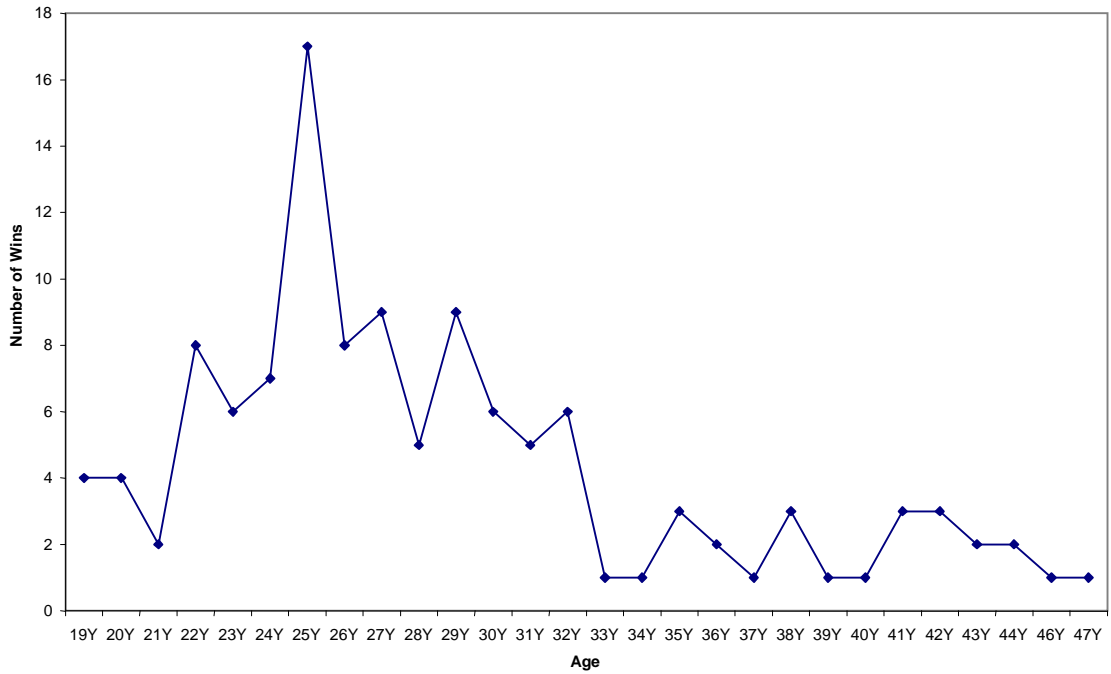

Fig. 4d: LPGA Tour Winning Trend (2002-2006, A. Sorenstam excluded)

## Regression Analyses

I also ran regression analyses to test the impact of age on winning golf games in each tour. Given that the trend between age and winning tours seems to have an inverted-U shape or curvilinear, models that combine linear and polynomial specification of age would best describe the relationship between the two variables. And if, indeed, the shape of the winning golf games takes an inverted-U shape, the slope for the polynomial-age variable will have a negative sign. This relationship can be shown as in Eq. 1.

$$Y = a + b_1 (X)_1 - b_2 (X^2)_2 + e \qquad (Eq.\ 1)$$

where Y = winning golf tournaments; a = the y-intercept; $b_1$ = the slope of the linear-age variable; $b_2$ = the slope of the polynomial-age variable; $X_1$ = the linear-age variable; $X_2$ = the polynomial-age variable; and e is the error term.

For instance, for the U.S. PGA Tour, I obtained the following results:

$$Y = -64 + 4.4 X_1 - 0.06 (X^2)_2$$

Using the equation of the regression line for the PGA Tour data, I found that the age at which golfers peaked their wins was 35. This held true when Tigers Woods and Vijay Singh were included from and excluded in the analyses. We have to be careful not to make a lot out of the regression results since the sample sizes are less than 30 for each tour (this is because golfers who share the same birth year are combined in the analyses). Nevertheless, these small-size regression analyses showed that the linear- and the polynomial-age variables are statistically significant for the PGA Tour, European PGA Tour, and LPGA Tour (when Anika Sorenstam is included). The slopes also depicted signs as expected. On the other hand, the age and winning relationship for the Champions Tour and for the LPGA Tour (when Sorenstam is excluded) could be described only by using the inverse linear regression model. In sum, the findings in the regression models seemed to support the results in the description section of this study.

## Conclusions

This research has found that the ages at which the PGA, European, Champions, and LPGA golfers peak their wins are 35, 30, 52, and 25, respectively. The regression analyses I conducted seemed to support my hypothesis that age affects winning professional golf tournaments. It should be noted, however, that the purpose of this study was to describe and test the relationship between age and winning golf games. But other variables not controlled in this study could also influence winning golf tours. For instance, golfers' skills, motivations, physical fitness, and practice regiments will likely play major roles in the success of golfers and should be given greater import by analysts in future studies.